\documentclass[conference]{IEEEtran}
\IEEEoverridecommandlockouts
\pdfoutput=1
\usepackage{cite}
\usepackage{amsmath,amssymb,amsfonts}
\usepackage{algorithmic}
\usepackage{graphicx}
\usepackage{textcomp}
\usepackage{xcolor}
\usepackage{authblk}
\usepackage{mathtools}
\def\BibTeX{{\rm B\kern-.05em{\sc i\kern-.025em b}\kern-.08em
    T\kern-.1667em\lower.7ex\hbox{E}\kern-.125emX}}
\begin{document}

\title{Comprehensive Study on Denoising of Medical Images Utilizing Neural Network Based Auto-Encoder}
%\thanks{}

\author{Thoshara Nawarathne\textsuperscript{1}}
\author{Thanushi Withanage\textsuperscript{1}}
\author{Samitha Gunarathne\textsuperscript{1}}
\author{\\Upekha Delay\textsuperscript{1}}
\author{Eranda Somathilake\textsuperscript{2}}
\author{Janith Senanayake\textsuperscript{1}}
\author{\\Roshan Godaliyadda\textsuperscript{1}}
\author{Parakrama Ekanayake\textsuperscript{1}}
\author{Janaka Wijayakulasooriya\textsuperscript{1}}
\author{\\Chathura Rathnayake\textsuperscript{3}}

\affil{\textsuperscript{1}Department of Electrical and Electronic Engineering, University of Peradeniya, Sri Lanaka.}\affil{\textsuperscript{2}Department of Mechanical Engineering, University of Peradeniya, Sri Lanka.}\affil{\textsuperscript{3}Department of Obstetrics and Gynaecology, University of Peradeniya, Sri Lanka.\\

\authorcr  {\tt \{thoshara.nawarathne, thanushiwithanage, samithalg\}@eng.pdn.ac.lk\textsuperscript{1}}
\authorcr{ \tt \{upekha.delay\textsuperscript{1}, eranda.somathilake\textsuperscript{2}, janith.b.senanayaka\textsuperscript{1}\}@eng.pdn.ac.lk} 
\authorcr  {\tt \{roshangodd, mpb.ekanayake,jan\}@ee.eng.pdn.ac.lk\textsuperscript{1},chathura67@hotmail.com\textsuperscript{3}}\vspace{1.5ex}}

\maketitle

\begin{abstract}
Fetal motion discernment utilizing spectral images extracted from accelerometric data incident on pregnant mothers abdomen has gained substantial attention in the state-of-the-art research. It is an essential practice to avoid adverse scenarios such as stillbirths and intrauterine growth restrictions. However, this endeavor of ensuring fetus safety has been arduous due to the existence of random noise in medical images. This novel research is an in depth approach to analyze how the interference of different noise variations affect the retrieval of information in those images. For that, an algorithm employing auto-encoder-based deep learning was modeled and the accuracy of reconstruction of the STFT images mitigating the noise has been measured examining the loss. From the results, it is manifested that even a substantial addition of the Super-Gaussian noises which have a higher correlation of the frequencies possessed by the Fetal movement images can be restored successfully with the slightest error.
\end{abstract}

\begin{IEEEkeywords}
Fetal movement, Auto-encoder, Neural network, Denoising, STFT
\end{IEEEkeywords}

\section{Introduction}
Biomedical image processing is essential to extract the useful information embedded in most of the biomedical signals. These processed images are useful for doctors in identifying adverse health conditions \cite{b1}. One of the most current areas of interest in medicine is ensuring the fetus's health by discerning the activeness of the baby inside the womb to avoid miscarriage and intrauterine growth restrictions. Currently, there are several types of research going on to detect fetal movements from the spectral images constructed utilizing the signals obtained from the pregnant mother’s abdomen\cite{b2} \cite{b3} \cite{b4} \cite{b5}. However, due to the methods utilized to obtain these data, there's a tendency of the original signals being interfered by external noises such as the addition of mother’s movements, and the effect from the materials used in the wearable device. This makes the fetal movement differentiation problem extremely complex resulting in many false-positive cases. Hence the difficulty in analyzing fetal well-being arises. Therefore, denoising these raw images is extremely crucial as a preliminary step prior to information extraction \cite{b6}. 

There are many denoising algorithms in the literature such as the Non-Local Mean algorithm (NLM) \cite{b7} and Total Variation Regularization Algorithm \cite{b8}, but the most effective and recent method is Convolution Denoising Auto-Encoders (CDAE) which is make use of convolution deep auto-encoder method. There are several research attempts \cite{b9} \cite{b10} where they have used the auto-encoder approach to denoise medical images. However those attempts were confined to analyzing only some specific noise cases and focused on the structure of the neural network. Hence not much significant research have been conducted to comprehensively analyze the effect of all noise varieties in image denoising.

For this research first, we selected fetal signals with low noise. Initially the data were in the time domain. By utilizing short Time Fourier Transform, images were generated to notice both temporal and frequency domain variations \cite{b11}. Thereafter, in the denoising task, we used an auto-encoder and analyzed its performance in accurate de-nosing. Subsequently,
the noises with different noise factors were tested. Additionally, how the noise properties such as colour, kurtosis, and skewness \cite{b12} affect the denoising process of the images were observed. For this purpose, different noise distributions were appended to the original image, and reconstruction loss was tracked through an auto-encoder neural network.

\section{Objective}
Current fetal movement detection methods include MRI and CT scans. There is a risk that those may emit harmful radiations to the fetus \cite{b13}\cite{b14}. We found a data set that comprises accelerometric data obtained in a non-invasive manner \cite{b15}. But it was observed that those signals had the tendency to get interfered with external noises. In this research, the intention is to employ an auto-encoder to denoise the images and to utilize this proposed algorithm as a pre-processing step for the classification of the fetal movements.

\section{Data Set}
 
For this research, a Biomedical data set named "Fetal Movement Detection Dataset Recorded Using MPU9250 Tri-Axial Accelerometer" was utilized. In it, MPU 9250 tri-axial accelerometer data including gyration, external movements, mothers' responses have been recorded of 13 singleton mothers with an average gestation period of 28 to 40 weeks. Each session contains an average of 7 fetal movements. 

\section{Methodology}\label{AA}

This section will express the proposed scheme for biomedical image denoising. It consists of pre-processing steps and auto-encoder based image denoising methods as shown in Fig. 1.

\begin{figure}[htbp]
\centerline{\includegraphics [width=0.52\textwidth] {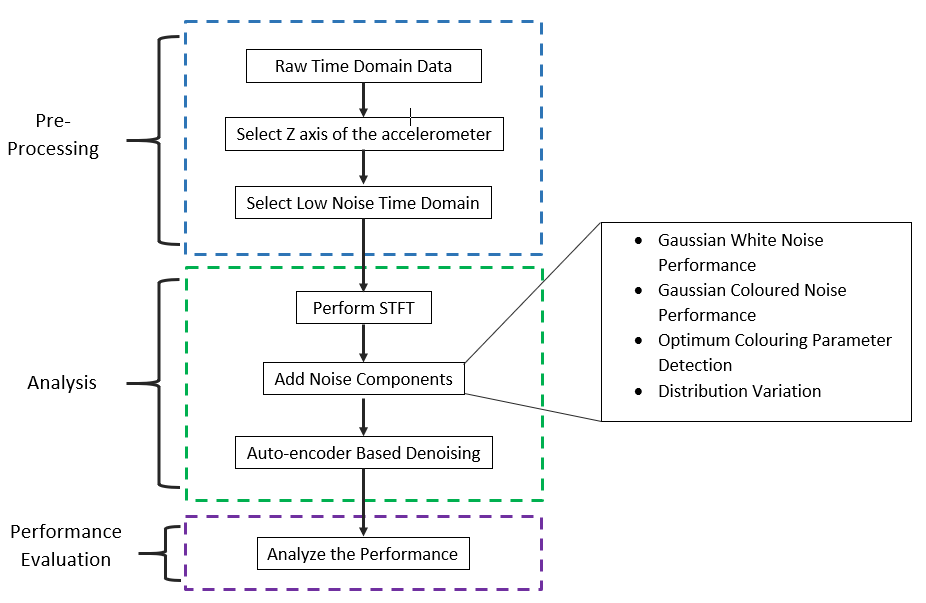}}
\caption{Denoising Procedure}
\label{fig}
\end{figure}

 The data of the previously mentioned data set belong to the time domain.
  It can be observed in Fig. 2 the time-domain interpretation of the fetal movement signals and the impact of the fetal movement is most dominant in the Z-axis of the accelerometer. Therefore, it was used for further denoising.
 
 \begin{figure}[htbp]
\centerline{\includegraphics[width=0.50\textwidth]{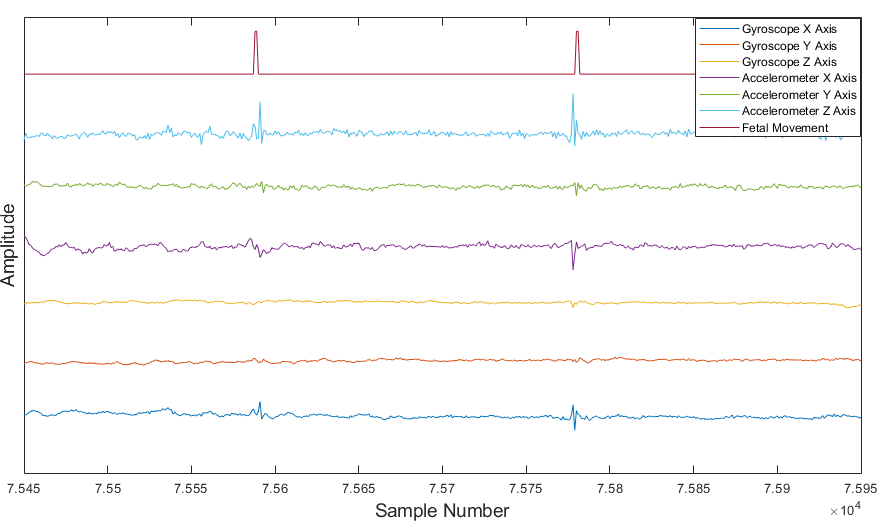}}
\caption{Fetal Movement Observation According to Various Axes of the Sensor }
\label{fig}
\end{figure}

Fig. 3 shows the low noise and high noise fetal movement signal represented in the Time Domain. 

\begin{figure}[htbp]
\centerline{\includegraphics[width=0.52\textwidth]{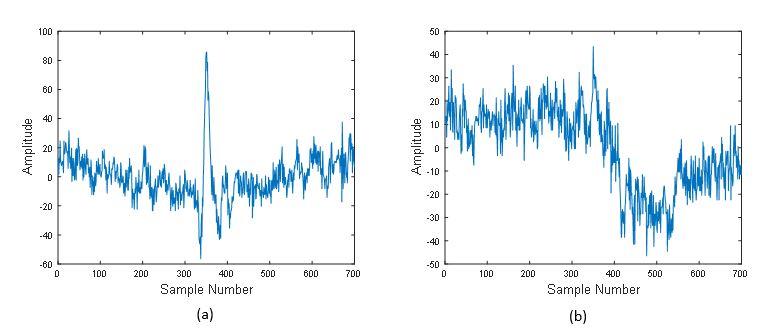}}
\caption{(a)Time Domain Low noise Image, (b)Time Domain Noisy image}
\label{fig}
\end{figure}

The selected low noise time-domain signals transformed into frequency-time variation by utilizing the Short Time Fourier Transform method.
 
\subsection{Short Time Fourier Transform} 
The Short-time Fourier transform (STFT) is a subcategory of Fourier transform. It is a particularly crucial section of the temporal - frequency distribution of a signal. STFT can be considered as a method of identifying the complex amplitude of a signal with respect to its time and frequency, as well as the phase\cite{b16}. 
 
From an analytical perspective, the STFT of a signal is computed in two steps. In the first stage, the total signal is divided into short time sections, where the length of each section takes as a unique value, which refers to as the window size. This window length should be selected accordingly to optimize the output\cite{b17}. In the second step, Fourier transform of each shorter sections were computed separately using equation(1). This explores the Fourier spectrum of each small segment of the longer signal. Eventually, the visual representation of the STFT is obtained, which interprets the segment-frequency content versus time. This output plot is named Spectrogram.

\begin{equation}
 F(t', u) = \sum_{t=-\infty}^{+\infty} f(t).W(t - t').e^{-j2\pi ut}
\end{equation}

where,\par
    \makebox[1.5cm]{\(F(t', u)\)} - STFT of \(f(t)\) :computed for each window\par 
    \makebox[1.5cm]{\(\ \)}\hspace{0.25cm} centered at \(t = t'\)\par
    \makebox[1.5cm]{\(\ t'\)} - Time parameter\par
    \makebox[1.5cm]{\(u\)} - Frequency parameter\par
    \makebox[1.5cm]{\(f(t)\)} - Signal to be analyzed\par
    \makebox[1.5cm]{\(W(t - t')\)} - Windowing function centered at \(t = t'\)\par
\bigskip

MATLAB was used to generate the STFT images for the above data. For this purpose, initially, the raw time-domain signal was segmented by selecting the sample size as 300, considering the average time elapse of the fetal movement signal. Subsequently, the windowing method, window size, and overlap for the spectrogram were selected adequately. The transformed sample spectrograms for both low noisy and high noisy signals in Fig. 3 can be observed in Fig. 4.

\begin{figure}[htbp]
\centerline{\includegraphics[width=0.5\textwidth]{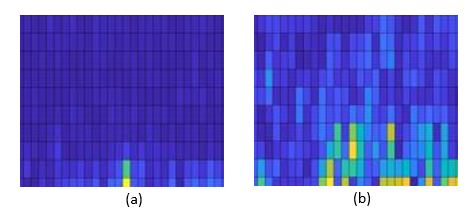}}
\caption{(a)Spectrogram Low Noise Image, (b)Spectrogram Noisy Image}
\label{fig}
\end{figure}

Then, the generated STFT images were resized and set as the inputs to the auto-encoder.

\subsection{Auto Encoders}

An auto-encoder is an unsupervised deep artificial neural network, which can be effectively utilized for data coding. It is equipped with propagation and back propagation.

Learning of a representation of a particular set of data is the fundamental purpose of an auto-encoder. It consists of both encoder and decoder. The encoder side is particularly known as the reduction side.  Specifically, dimension reduction is achieved by training the network to filter the noise of the original signal. This can be achieved through convolutional bottleneck architecture\cite{b18}. On the other hand, on the decoding side, the auto-encoder tries to generate the reduced encoded output, that merely matches with the original input signal. 
In practice, there are two basic variations of auto-encoders: Regularized Auot-encoders and Variational Auto-encoders. Under the regularized auto-encoder sector it can be further categorized as Sparse auto-encoders(SAE), Denoising auto-encoders (DAE), and Contractive auto-encoders(CAE). This paper is based on the denoising auto-encoders.

Denoising auto-encoder focuses on the reduction of noise through identity mapping. To use the auto-encoder to achieve the above-mentioned goal, it should first train the network \cite{b19}. 

In this process, initially, the no noise (low noise) images were applied as inputs ($X$). Then, noises were added to create noisy images ($X'$). After feeding through the denoising auto-encoder, the images can be reconstructed such that minimum noise was obtained. Encoding algorithm is represented in equation (2).

\begin{equation}
Encoder: Y = A(WX'+b)
\end{equation}

$X'$ images were taken by the encoder and it constructed the output $Y$. Here, $A$ represents the activation function, and $W$ is the weight added to the hidden layer(s). Besides, a bias $b$ is also added in there.

On the decoder side, the latent representation of $Y$ was mapped again to reconstruct the images $Z$, which were assumed to be low noisy. The decoder was modeled as stated in equation (3).

\begin{equation}
Decoder: Z = A(W'Y+b')
\end{equation}

Where $W'$ and $b'$ are the weights and the biases added in the decoder respectively. 

Here the auto-encoder parameters such as $W$, $W'$, $b$, and $b'$ were tuned to minimize the cost function. It was achieved by varying the parameters such as the number of layers, number of neurons per layer, and neuron size.  

The performance of denoising auto-encoder was analyzed by varying the fraction of noise added and characteristics of the noise: whiteness and distribution etc.

\begin{figure}[htbp]
\centerline{\includegraphics [width=0.47\textwidth] {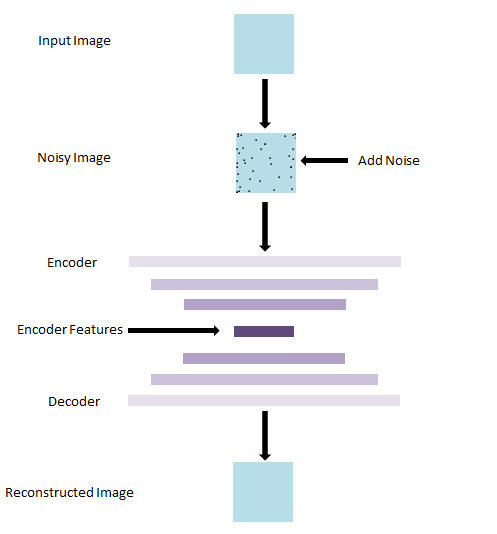}}
\caption{Auto-encoder Architecture of Denoising Images}
\label{fig}
\end{figure}

This encoder-decoder denoiser architecture can be observed in Fig. 5.  To feed the network, initially, 75 low noisy fetal movement signals were selected. After transforming them into spectrograms those were resized to create images with size   \(256\times256\times3\). Then the noise addition was done in four separate sections as listed below.

\begin{enumerate}
\item Gaussian White Noise Performance
\item Gaussian Coloured Noise Performance
\item Optimum Colouring Parameter Detection
\item Distribution Variation
\end{enumerate}

The used auto-encoder architecture was a sequential model, which was analyzed using Tensor Flow libraries. This model consisted of two convolutional layers for the encoder and two for the decoder. In the encoder section, the first convolutional layer used 32 latent variables with a kernel size of \(3\times3\), with Relu activation. This was followed by the max-pooling layer with a window size of \(2\times2\). The next hidden layer had 64 neurons, kernel size of \(3\times3\), with a Relu activation and a max-pooling layer with the same dimensions as above. Decoder had the parameters with 64 and 32 feature maps, \(3\times3\) filter size, with \(2\times2\) max-pooling layers activated by Relu function. 

The performance evaluation according to the different noise types and parameters is discussed in further sections.
\\

\subsubsection{\textbf{Gaussian White Noise Performance}}

To analyze the auto-encoder denoising performance concerning different characteristics of added noise, initially, a Gaussian-white noise signal was added. The addition of Gaussian white noise signal represents an ideal situation when the image gets interfered with noises at every frequency in an equal power level and the amplitude distribution of those noises represents a normal distribution \cite {b20}. Here, the amount of noise addition was varied by changing the Noise Factor parameter from 0 to 0.9 by 0.1 intervals, where a Noise Factor of 0.9 means the images are almost completely noisy. This relationship is shown in equation (4). The objective of this section was to identify the optimum noise factor such that it results in the minimum loss of the algorithm.

\begin{equation}
Noise\hspace{1mm} Factor=\dfrac{Noise\hspace{1mm} Power}{Signal\hspace{1mm} Power}
%{eq}
\end{equation}
\
\subsubsection{\textbf{Gaussian Coloured Noise Performance}}\label{SCM}

The nature of the input fetal movement signals is not white. Therefore, it is better to explore the denoising algorithm to coloured noise components. To achieve that goal colouring of the Gaussian-white signal was done using applying the filter depicted in Fig. 6. By applying that filter transformation to white Gaussian noise, some noise frequency components are suppressed, while changing the overall shape of the power spectral density of the Gaussian noise distribution.
\begin{figure}[htbp]
\centerline{\includegraphics [width=0.47\textwidth] {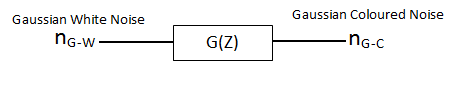}}
\caption{Filter Architecture Used for Colouring Noise}
\label{fig}
\end{figure}
\\
\begin{equation}
G(Z)=\dfrac{\sqrt{1-a^2}}{1-aZ^{-1}}
%{eq}
\end{equation}
\
\subsubsection{\textbf{Optimum Colouring Parameter Detection}}

Next, our approach was to detect the optimum alpha in which the proposed auto-encoder network presents the highest accuracy in reconstructing the noisy spectral fetal movement images. So that such types of noises included in the medical images could be effectively denoised. For that, the coloring parameter, `$a$' in equation (5) was changed from 0 to 0.9 in steps of 0.1 to generate Gaussian signals of different color levels. These signals are shown in Fig. 7. The minimum loss was examined considering both various '$a$' values and noise factors.

\begin{figure}[htbp]
\centerline{\includegraphics [width=0.47\textwidth] {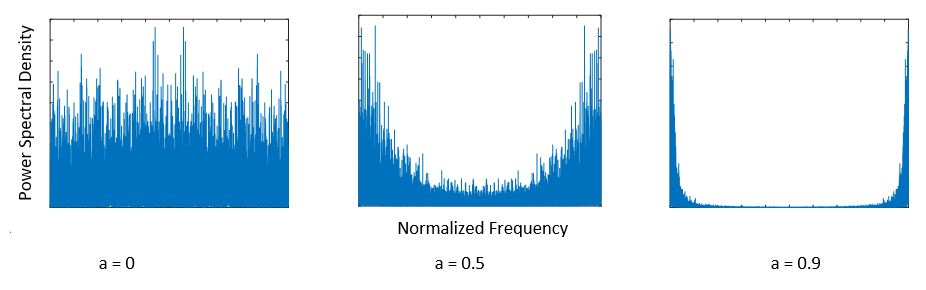}}
\caption{Generated Gaussian Noise for Different Colour Levels}
\label{fig}
\end{figure}

\subsubsection{\textbf{Distribution variation}}

Up to now, the above analysis was done considering only normally distributed signals. However, the added noise for bio-medical signals may be differing from Gaussian signals. Therefore, it is necessary to explore the behavior of the auto-encoder denoiser when the input images get added with non-Gaussian noise. Those non-Gaussian signals are further categorized as Sub-Gaussian and Super-Gaussian signals, which were generated by changing the features such as kurtosis and skewness of the noise distributions\cite{b21}. Hence, some standard distributions of varied kurtosis and skewness were used for this purpose. Uniform, Raised Cosine, and Weiner Semi-Circle distributions were used for exploring the Sub-Gaussianity nature whereas double exponential, Weibull, Hyperbolic Secant, Laplace, and Logistic distributions were used to analyze Super-Gaussian nature. Fig. 8 and Fig. 9 depicts the selected Super-Gaussian and Sub-Gaussian signal distributions respectively.

\begin{figure}[htbp]
\centerline{\includegraphics [width=0.5\textwidth] {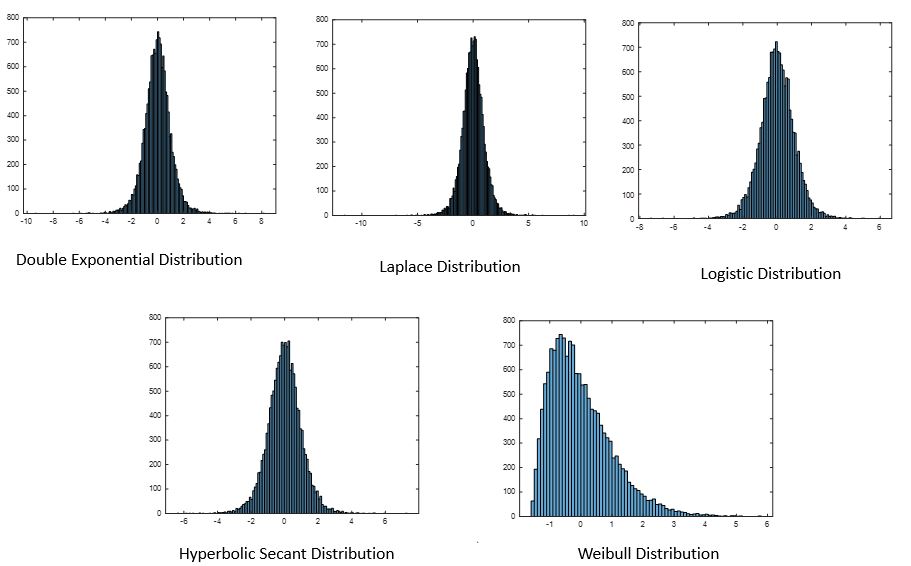}}
\caption{Selected Super-Gaussian Distributions}
\label{fig}
\end{figure}

\begin{figure}[htbp]
\centerline{\includegraphics [width=0.5\textwidth] {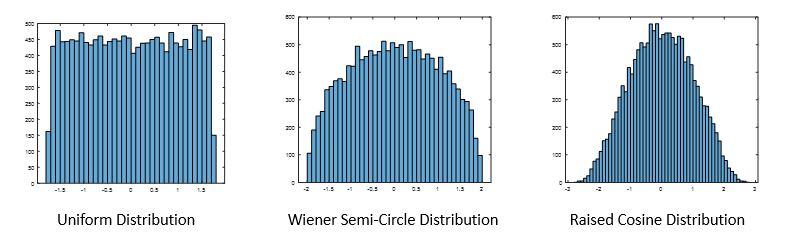}}
\caption{Selected Sub-Gaussian Distributions}
\label{fig}
\end{figure}

For the above distributions, parameters such as the noise factor and the coloring parameter were tuned. By analyzing the minimum loss in denoising, the optimum noise value and coloured factor, for each above mentioned distribution, were obtained.

\section{Results}

The minimum loss of the algorithm was varied with the noise factor when the White Gaussian noise was added to the input. Initially we try to optimize the minimum loss by varying Noise Factor as well as Number of Epochs. From this analysis, we obtained that there is a significant drop of loss around Noise Factor = 0.3 for 10 to 50 Epochs. This variation is depicted in Fig. 10.

\begin{figure}[htbp]
\centerline{\includegraphics [width=0.5\textwidth] {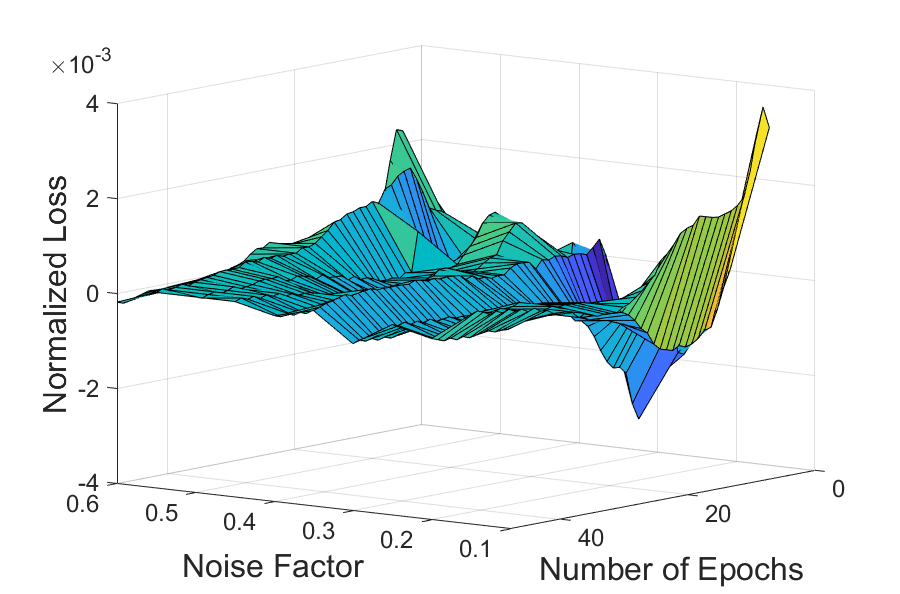}}
\caption{Variation of Normalized Loss with Noise Factor and Number of Epochs}
\label{fig}
\end{figure}

Considering the computational cost we selected to use the Number of Epochs as 30. For Number of Epochs = 30 acheivable minimum loss with respective to Noise Factor is shown in Fig. 11. As it exhibits there, the finest denoising performance can be obtained when the N.F. is about 0.3 for white Gaussian noise.

\begin{figure}[htbp]
\centerline{\includegraphics [width=0.5\textwidth] 
{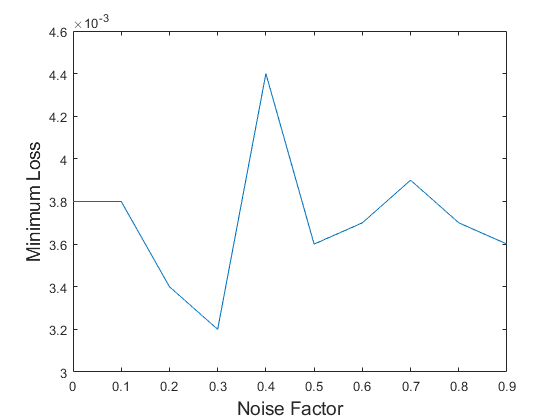}}
\caption{Variation of Minimum Loss vs. Noise Factor for White Gaussian Noise}
\label{fig}
\end{figure}

Then as the analysis extended to checking the performance in colored gaussian signals the same procedure was carried out by varying the N.F. for different '$a$' values. The observed variations are shown in Fig. 12. 

\begin{figure}[htbp]
\centerline{\includegraphics [width=0.5\textwidth] {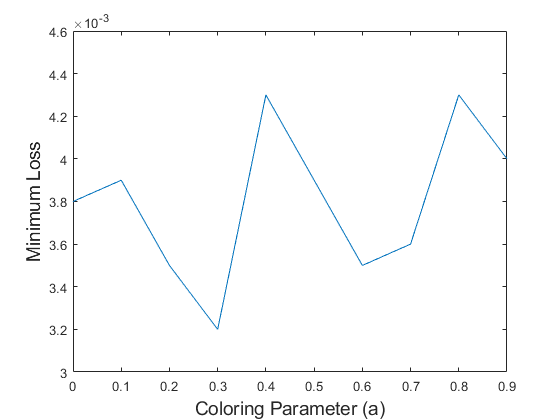}}
\caption{Variation of Loss vs. Colouring for Gaussian Noise for N.F. = 0.3}
\label{fig}
\end{figure}

\begin{figure}[htbp]
\centerline{\includegraphics [width=0.48\textwidth] {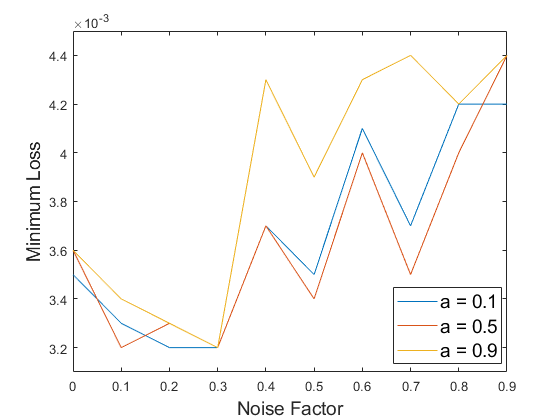}}
\caption{Variation of Loss vs. Noise Factor for Different '$a$' Values}
\label{fig}
\end{figure}

In further analysis, it is manifested that, here regardless of the value of the coloring parameter of the noise signal, the optimum noise factor was also obtained around 0.3. 
The obtained minimum loss variation with the noise factor for sample `$a$' values can be observed in Fig. 13.

 The obtained minimum loss was varied when varying the distribution of the noise. When comparing the Sub-Gaussian and Super-Gaussian white signals, Super-Gaussian signals perform well in the denoising of the biomedical images.

 Even though the added noise proportion was quite high, most of the Super-Gaussian noise components can be eliminated in a fairly accurate manner than Sub-Gaussian noise. These results are tabulated in Table I and sample behaviour for sub and super Gaussian distribution is depicted in Fig. 14. 

\begin{table}[htbp]
\caption{Noise Factors, Which Obtained Minimum Loss (0.0032) of White Gaussian Noises with Different Probability Distributions}
\begin{center}
\begin{tabular}{|c|c|c|c|}
\hline
\textbf{Gaussianity}&\textbf{Distribution}&\textbf{Noise }&\textbf{Next Minimum}\\

\textbf{}&\textbf{}&\textbf{ Factor}&\textbf{Loss with N.F.}\\
\cline{1-4} 
\text{Sub-Gaussian}&\text{Raised Cosine}&\text{0.1}&\text{0.0034 at N.F. = 0.3}\\
\hline
\text{Sub-Gaussian}&\text{Uniform}&\text{0.2, 0.3}&\text{0.0033 at N.F. = 0.5}\\
\hline
\text{Sub-Gaussian}&\text{Wiener Semi-circle }&\text{0.1, 0.3}&\text{0.0034 at N.F. = 0.4}\\
\hline
\text{Super-Gaussian}&\text{Double Exponential}&\text{0.2, 0.3}&\text{0.0034 at N.F. = 0.8}\\
\hline
\text{Super-Gaussian}&\text{Hyperbolic Secant}&\text{0.2, 0.4}&\text{0.0033 at N.F. = 0.6}\\
\hline
\text{Super-Gaussian}&\text{Laplace}&\text{0.2, 0.3}&\text{0.0034 at N.F. = 0.8}\\
\hline
\text{Super-Gaussian}&\text{Logistic}&\text{0.3, 0.4}&\text{0.0033 at N.F. = 0.5}\\
\hline
\text{Super-Gaussian}&\text{Weibull}&\text{0.2, 0.5}&\text{0.0034 at N.F. = 0.7}\\
\hline
%multicolumn{4}{l}{$^{\mathrm{a}}$Sample of a Table footnote.}
\end{tabular}
\label{tab1}
\end{center}
\end{table}

\begin{figure}[htbp]
\centerline{\includegraphics[width=0.52\textwidth]{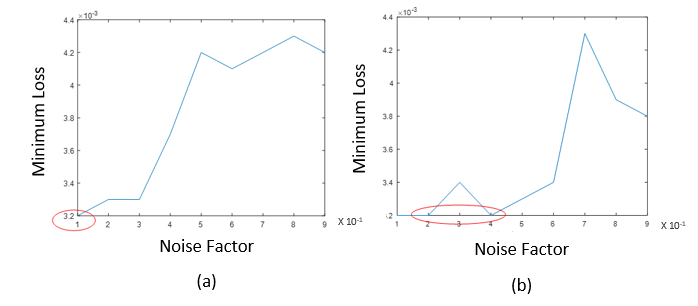}}
\caption{(a)Minimum Loss for Sub-Gaussan Noise, (b)Minimum Loss for Super-Gaussan Noise}
\label{fig}
\end{figure}

If the input noise is colored, and its distribution was varied the obtained minimum loss also varies. Their performances are shown in Table II. 

\begin{table}[htbp]
\caption{Minimum Average Loss for Different Distributions with Optimum Colouring Parameter}
\begin{center}
\begin{tabular}{|c|c|c|c|}
\hline
\textbf{Gaussianity}&\textbf{Distribution}&\textbf{Colouring }&\textbf{Average }\\

\textbf{}&\textbf{}&\textbf{ Parameter}&\textbf{Minimum Loss }\\
\cline{1-4} 
\text{Sub-Gaussian}&\text{Raised Cosine}&\text{0.5}&\text{0.0036}\\
\hline
\text{Sub-Gaussian}&\text{Uniform}&\text{ 0.7}&\text{0.0038}\\
\hline
\text{Sub-Gaussian}&\text{Wiener Semi-circle }&\text{0.5}&\text{0.0037}\\
\hline
\text{Super-Gaussian}&\text{Double Exponential}&\text{0.4}&\text{0.0036}\\
\hline
\text{Super-Gaussian}&\text{Hyperbolic Secant}&\text{0.8}&\text{0.0034}\\
\hline
\text{Super-Gaussian}&\text{Laplace}&\text{0.7}&\text{0.0038}\\
\hline
\text{Super-Gaussian}&\text{Logistic}&\text{0.7}&\text{0.0036}\\
\hline
\text{Super-Gaussian}&\text{Weibull}&\text{0.6}&\text{0.0036}\\
\hline
%multicolumn{4}{l}{$^{\mathrm{a}}$Sample of a Table footnote.}
\end{tabular}
\label{tab1}
\end{center}
\end{table}

 Also, for a sample Gaussian noise with sub-gaussian distribution the loss variation is shown in Fig. 15, whereas Fig. 16 depicts the loss varying behavior of a sample Super-Gaussian distributed Gaussian signal. 
\begin{figure}[htbp]
\centerline{\includegraphics [width=0.51\textwidth] {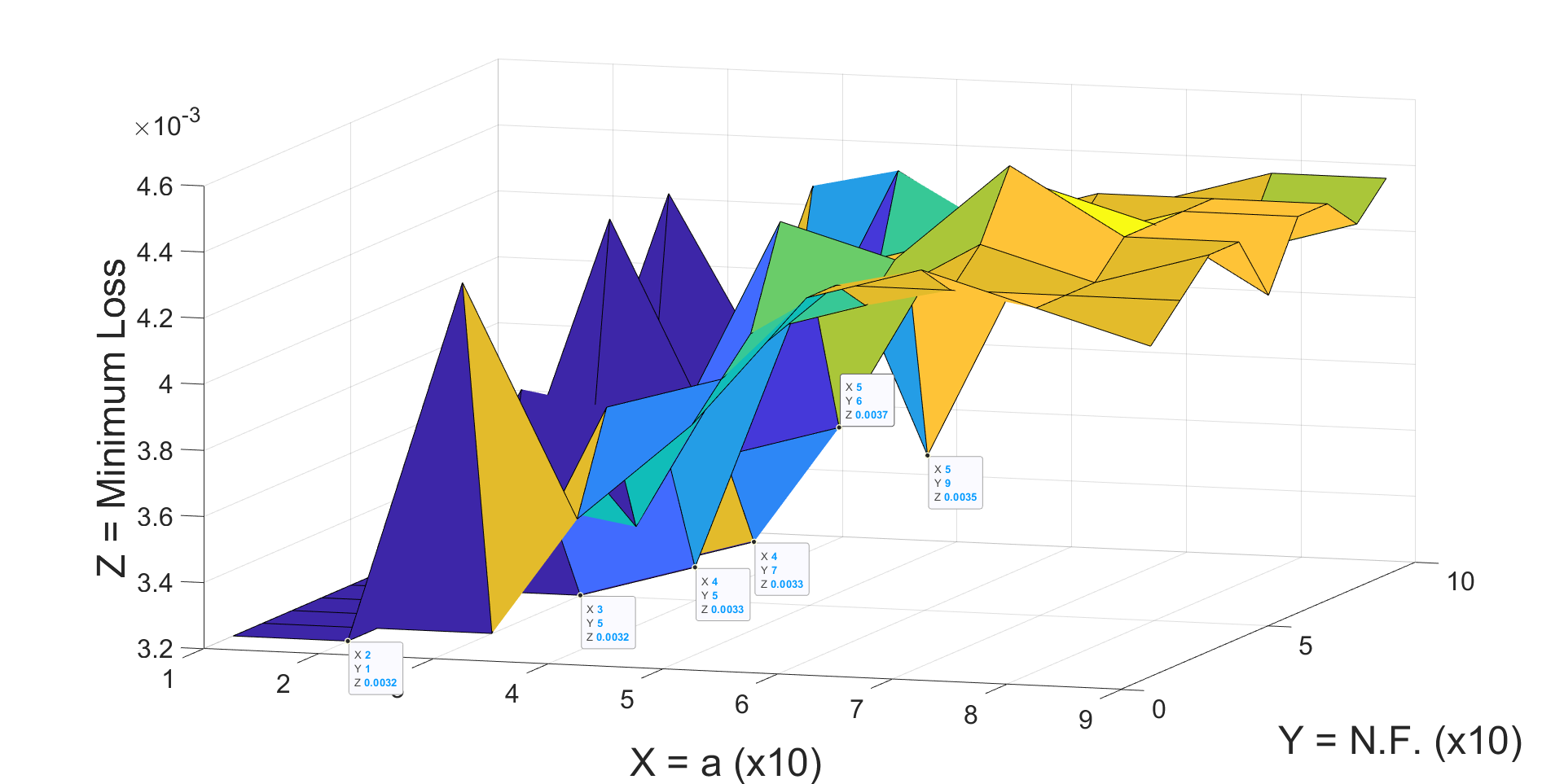}}
\caption{{Variation of Minimum Loss and N.F. vs. Colouring Parameter for a Noise with Uniform Distribution}}
\label{fig}
\end{figure}

\begin{figure}[htbp]
\centerline{\includegraphics [width=0.51\textwidth] {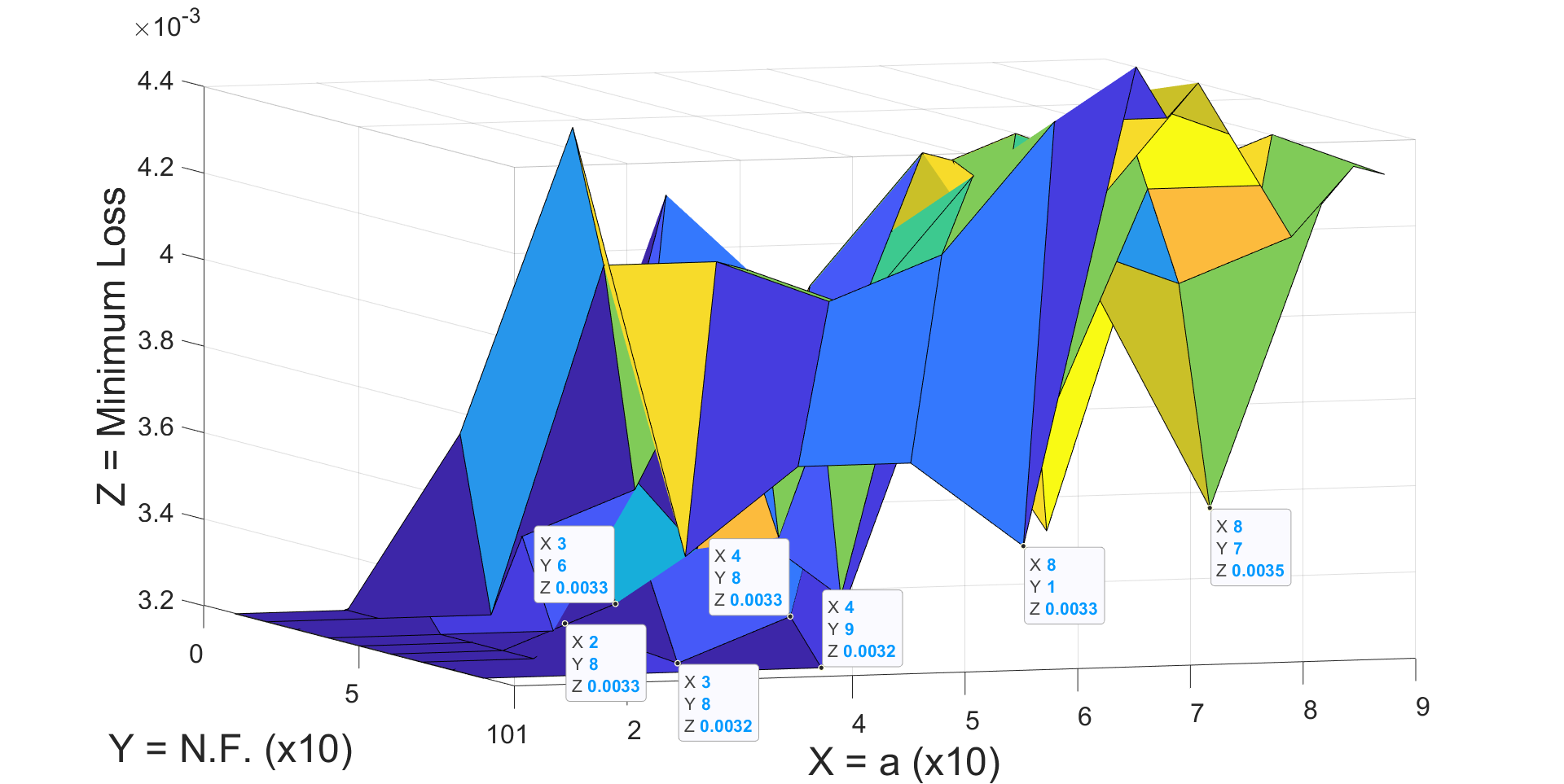}}
\caption{Variation of Minimum Loss and N.F. vs. Colouring Parameter for a Noise with Double Exponential Distribution}
\label{fig}
\end{figure}

\section{Conclusion}

Image denoising is crucial when analyzing the bio-medical images. Here to identify the fetal movement signals with considerable accuracy it is better to use a denoising method as a pre-processing step before feeding into a classifier. In this research denoising method based on convolutional auto-encoders was applied to spectral images, and its performance was evaluated. From the above analysis, it can be observed that when Gaussian-White or Gaussian-coloured noise was added the optimum denoising was obtained when Noise Factor is 0.3. Further, the optimum coloring parameter of the denoiser was alpha = 0.3. 

Thereafter, varying the noise distribution from Gaussian to Sub/Super-Gaussian noise signals with optimum noise factor varied gives promising results. Specifically for Super-Gaussian signals, even the noise factor more than 0.7-0.8 offers the finest denoising performances displaying high accuracy for both white and coloured noise components.  This may result since most of the noise signals in nature vary in a smooth manner. So the  fetal signals  impacted by these soft variations can be modeled using the Super-Gaussian signals. Therefore, to denoise the fetal movement biomedical signals it is ideal to use the above proposed convolutional auto-encoder denoiser.

\vspace{12pt}

\end{document}